\documentclass[12pt]{article}
\usepackage[dvips]{graphicx}
\usepackage{amssymb}
\usepackage{amsmath,amsthm}

\usepackage{bm}
\bmdefine{\Bzero}{0}
\bmdefine{\Bone}{1}

\def\Bx{{\bf x}}
\def\By{{\bf y}}
\def\Bz{{\bf z}}

\def\Bbeta{{\bf \beta}}

\def\BA{{\rm A}}
\def\BB{{\rm B}}
\def\BC{{\rm C}}
\def\BD{{\rm D}}
\def\BE{{\rm E}}
\def\BF{{\rm F}}
\def\BG{{\rm G}}

\def\BI{{\rm I}}

\newtheorem{remark}{Remark}[section]

\newtheorem{fact}{Fact}[section]
\title{Markov basis for design of experiments with three-level factors}
\author{
Satoshi Aoki\\
Department of Mathematics and Computer Science\\
Kagoshima University\\
and\\
Akimichi Takemura\\
Graduate School of Information Science and Technology\\
University of Tokyo
}
\date{February, 2008}
\begin{document}
\maketitle
\begin{abstract}
We consider Markov basis arising from fractional
 factorial designs with three-level factors. 
 Once we have a Markov basis, $p$ values for various
 conditional tests are estimated by the Markov chain Monte Carlo
 procedure. For designed experiments with a single count observation for
 each run, we formulate a generalized linear model and consider a sample
 space with the same sufficient statistics to the observed data. 
 Each model is characterized by a covariate matrix, which is
 constructed from the main and the interaction effects
 we intend to measure. We investigate fractional factorial designs with
 $3^{p-q}$ runs noting correspondences to the models for $3^{p-q}$
 contingency tables. 
\end{abstract}

\section{Introduction}
In the past decade, a new application of computational algebraic
techniques to statistics has been developed rapidly.  Diaconis and Sturmfels
\cite{DS98} introduced the notion of {\it Markov basis} and presented
a procedure for sampling from discrete conditional distributions by
constructing a connected, aperiodic and reversible Markov chain on a 
given sample space.  Since then, many works have been published on the
topic of the Markov basis
by both algebraists and statisticians.  
Contributions of the present authors on Markov bases can be found in
\cite{AHOT07}, \cite{AT03}, \cite{AT05a}, 
\cite{AT06}, \cite{AT07}, \cite{AT05b}, 
\cite{ATY05}, \cite{HAT06}, \cite{TA04} and \cite{TA05}. 
On the other hand, series of works by
Pistone and his collaborators (e.g., \cite{PW96}, \cite{RR98},
\cite{PRW00}, \cite{GPR03} and \cite{PR07}) successfully applied the theory of
Gr\"obner basis to designed experiments. 
In these works, a design is
represented as the variety for a set of polynomial equations.

In view of these two main areas of algebraic statistics, it is of
interest to investigate statistical problems which are related to both
designed experiments and the Markov basis.  In \cite{AT06} we
initiated the study of conditional tests of the main effects and the
interaction effects when count data are observed from a designed
experiment.  In \cite{AT06} we investigated Markov bases arising from
fractional
factorial designs with two-level factors.  In this paper, extending
the results in our previous paper, we consider Markov bases for 
fractional factorial designs with three-level factors.   
Motivated by comments by a referee, we also 
discuss relations between the Markov basis approach and  the 
Gr\"obner basis approach to designed experiments, 
although the connection between them are not yet very well
developed.  In considering alias relations for fractional
factorial designs, we mainly
use a classical notation, as explained in
standard textbooks on designed experiments such as \cite{WH00}.  
We think that the classical notation is more accessible to 
practitioners of experimental designs and 
our proposed method is useful for practical applications.
However, mathematically the aliasing relations can be more elegantly 
expressed in the framework of algebraic statistics by Pistone et al.
We make this connection clear in remarks in Section 2.

We relate the models for the case of fractional factorial designs to
various models of contingency tables.  In most of the works on Markov
bases for contingency tables, the models considered are hierarchical
models.  On the other hand, when we map models for fractional factorial
designs to models for contingency tables, the resulting models are not
necessarily hierarchical. Therefore Markov bases 
for the case of fractional factorial designs often have different
features than Markov bases for hierarchical models.
In particular for the fractional factorial designs with three-level
factors, we find interesting degree three moves and 
indispensable fibers with three elements.
These are of interest also from the algebraic viewpoint.

The construction of this paper is as follows. In Section 2, we
illustrate the problem of this paper and describe the testing
procedure for evaluating $p$ values of the main and the interaction
effects for controllable factors in designed experiments. 
Similarly to the preceding works on Markov basis for contingency
tables, our approach is to construct a connected Markov
chain for some conditional sample space. 
We explain how to define this sample space corresponding to various
null hypotheses.
In Section 3, we consider the relation
between the models for the
contingency tables and 
the models for the designed experiments for fractional factorial
designs with three-level factors.
Then we state properties of Markov bases for designs which are
practically important.
In Section 4, we give some discussion.

\section{Markov chain Monte Carlo tests for designed experiments}
In this section we illustrate the problem of this 
paper. We consider the Markov chain Monte Carlo procedure for
conditional tests of the main and the interaction effects of controllable
factors for the discrete
observation derived from various designed experiments. 
Our arguments are based on
the theory of the generalized linear models (\cite{MN89}). 

\subsection{Conditional tests for discrete observations}
Suppose that the observations are counts of some events
and one observation is obtained 
for each run of a designed experiment, which is defined by some
{\it aliasing relation}.  (We also consider the case that the observations
are the ratio of counts in Section 4.) 
For example, Table 1 is a $1/8$ fraction of a full factorial design 
(i.e., a $2^{7-3}$ fractional factorial design) defined from the 
aliasing relation
\begin{equation}
\BA\BB\BD\BE = \BA\BC\BD\BF = \BB\BC\BD\BG = \BI.
\label{eqn:aliasing-relation-tab1}
\end{equation}
This data set was considered in \cite{AT06} 
with some modification from the original data 
in \cite{C93}.  The original data was reanalyzed in \cite{HN97}. 
\begin{table*}
\begin{center}
\caption{Design and number of defects $y$ for the wave-solder experiment}
\begin{tabular}{ccccccccr}\hline
& \multicolumn{7}{c}{Factor} & \multicolumn{1}{c}{$y$}\\
Run & A & B & C & D & E & F & G & \\ \hline
 1 & 0 & 0 & 0 & 0 & 0 & 0 & 0 & 69\\
 2 & 0 & 0 & 0 & 1 & 1 & 1 & 1 & 31\\
 3 & 0 & 0 & 1 & 0 & 0 & 1 & 1 & 55\\
 4 & 0 & 0 & 1 & 1 & 1 & 0 & 0 & 149\\
 5 & 0 & 1 & 0 & 0 & 1 & 0 & 1 & 46\\
 6 & 0 & 1 & 0 & 1 & 0 & 1 & 0 & 43\\
 7 & 0 & 1 & 1 & 0 & 1 & 1 & 0 & 118\\
 8 & 0 & 1 & 1 & 1 & 0 & 0 & 1 & 30\\
 9 & 1 & 0 & 0 & 0 & 1 & 1 & 0 & 43\\
10 & 1 & 0 & 0 & 1 & 0 & 0 & 1 & 45\\
11 & 1 & 0 & 1 & 0 & 1 & 0 & 1 & 71\\
12 & 1 & 0 & 1 & 1 & 0 & 1 & 0 &380\\
13 & 1 & 1 & 0 & 0 & 0 & 1 & 1 & 37\\
14 & 1 & 1 & 0 & 1 & 1 & 0 & 0 & 36\\
15 & 1 & 1 & 1 & 0 & 0 & 0 & 0 &212\\
16 & 1 & 1 & 1 & 1 & 1 & 1 & 1 & 52\\ \hline
\end{tabular}
\end{center}
\end{table*}
The observation $y$ in Table 1 is the number of defects found in 
wave-soldering process in attaching components to an electronic circuit
card. In Chapter 7 of \cite{C93}, seven factors of a
wave-soldering process are considered: (A) prebake condition, (B) flux density, (C)
conveyer speed, (D) preheat condition, (E) cooling time, (F) ultrasonic
solder agitator and (G) solder temperature.
Each factor of Table 1 has two-level, which we write $0$ or $1$ in this
paper. 
The aim of this experiment is to decide which levels for each factor
are desirable to reduce solder defects.

\begin{remark}
\label{rem:2factor-alias}
  Specification and notation of aliasing relations in
  (\ref{eqn:aliasing-relation-tab1}) are explained in standard
  textbooks on designed experiments (e.g.\ \cite{WH00}) and well
  understood by practitioners of designed experiments.  As explained
  in Section 1.3 and Section 4.6 of \cite{PRW00}, the aliasing
  relations are more elegantly expressed as a set of polynomials
  defining an ideal in a polynomial ring.  Consider $\BA, \BB, \dots,
  \BG$ as indeterminates and let ${\mathbb C}(\BA, \BB, \dots, \BG)$
  the ring of polynomials in $\BA, \BB, \dots, \BG$ with complex coefficients.
  Then the ideal 
\begin{equation}
\label{eq:alias-ideal}
\langle \BA^2-1, \BB^2-1, \dots, \BG^2 -1, 
\BA\BB\BD\BE -1, \BA\BC\BD\BF-1, \BB\BC\BD\BG -1\rangle
\end{equation}
determines the aliasing relations, i.e., two interaction effects are
aliased with each other if and only if their difference belongs to the
ideal (\ref{eq:alias-ideal}). Given a particular term order, 
the set of standard monomials corresponds to a particular choice of
saturated model, which can be estimated from the experiment.
\end{remark}

Extending the above setting, 
in this paper, we consider three-level designs with count
observations. For example, Table 2 shows a $3^{4-2}$ fractional
factorial design and the observations.
\begin{table*}
\begin{center}
\caption{Design and observations for a $3^{4-2}$ fractional factorial design}
\begin{tabular}{cccccr}\hline
& \multicolumn{4}{c}{Factor} & \multicolumn{1}{c}{$y$}\\
Run & A & B & C & D \\ \hline
1 & 0 & 0 & 0 & 0 & $y_1$\\
2 & 0 & 1 & 1 & 2 & $y_2$\\
3 & 0 & 2 & 2 & 1 & $y_3$\\
4 & 1 & 0 & 1 & 1 & $y_4$\\
5 & 1 & 1 & 2 & 0 & $y_5$\\
6 & 1 & 2 & 0 & 2 & $y_6$\\
7 & 2 & 0 & 2 & 2 & $y_7$\\
8 & 2 & 1 & 0 & 1 & $y_8$\\
9 & 2 & 2 & 1 & 0 & $y_9$\\ \hline
\end{tabular}
\end{center}
\end{table*}
We write the three levels as $\{0,1,2\}$. 
Note that the design in Table 2 is also derived from an aliasing relation,
\begin{equation}
\BC = \BA\BB,\ \BD = \BA\BB^2.
\label{eqn:relation-3^4-2}
\end{equation}
We give a more detailed explanation of these aliasing relations in Section 2.2.

We adopt the theory of the {\it generalized linear models} (\cite{MN89})
as follows. 
For these types of count data, it is natural to consider the Poisson model.
Write the observations as $\By = (y_1,\ldots,y_k)'$, where $k$ is the
number of runs and $y_i$'s are
mutually independently distributed with the mean parameter $\mu_i = E(y_i), 
i = 1,\ldots,k$. 
We express the mean parameter  $\mu_i$ as
\[
g(\mu_i) = \beta_0 + \beta_1x_{i1} + \cdots + \beta_{\nu}x_{i\nu},
\]
where $g(\cdot)$ is the link function and $x_{i1},\ldots,x_{i\nu}$ are
the $\nu$ covariates defined in Section 2.2. 
The sufficient statistic is written as $\sum_{i=1}^{k} x_{ij}y_i, j = 1,\ldots,\nu$.
The canonical link for
the Poisson distribution is $g(\mu_i) = \log\mu_i$.
For later use, we write the $\nu$-dimensional
parameter $\beta$ and the covariate matrix $X$ as
\begin{equation}
 \Bbeta = (\beta_0,\beta_1,\ldots,\beta_{\nu-1})'
\label{eqn:parameter-beta}
\end{equation}
and
\begin{equation}
 X = 
\left(\begin{array}{cccc}
1 & x_{11} & \cdots & x_{1\nu-1}\\
\vdots & \vdots & \cdots & \vdots\\
1 & x_{k1} & \cdots & x_{k\nu-1}
\end{array}
\right) = 
\left(\begin{array}{cccc}
\Bone_k & \Bx_1 & \cdots & \Bx_{\nu-1}
\end{array}
\right) ,
\label{eqn:def-X}
\end{equation}
where $\Bone_k = (1,\ldots,1)'$ is the $k$-dimensional column
vector consisting of $1$'s.  Then the likelihood function is written as
\[
\begin{array}{ccl}
 \displaystyle\prod_{i=1}^{k}\frac{\mu_i^{y_i}}{y_i!}e^{-\mu_i}
& = & \left(\displaystyle\prod_{i=1}^{k}\frac{e^{-\mu_i}}{y_i!}
\right)\exp\left(\displaystyle\sum_{i=1}^{k}y_i\log\mu_i
\right)\\
& = & \left(\displaystyle\prod_{i=1}^{k}\frac{e^{-\mu_i}}{y_i!}
\right)\exp\left(\beta_0\Bone_k'\By + \displaystyle\sum_{j =
1}^{\nu-1}\beta_{j}\Bx_j'\By\right)\\
& = & \left(\displaystyle\prod_{i=1}^{k}\frac{e^{-\mu_i}}{y_i!}
\right)\exp\left(\Bbeta'X'\By\right),
\end{array}
\]
which implies that the sufficient statistic for $\Bbeta$ is $X'\By =
(\Bone_k'\By,\Bx_1'\By,\ldots,\Bx_{\nu-1}'\By)$.

To define a conditional test, we specify the {\it null model} and
the {\it alternative model} in terms of the parameter $\beta$. 
To avoid confusion, we express the free parameters under the
null model as the $\nu$-dimensional
parameter (\ref{eqn:parameter-beta}) in this paper. 
Alternative hypotheses are expressed in terms of additional parameters.
For example, in
the case of various goodness-of-fit tests, the alternative model is the
saturated model, i.e., $\beta$ is $k$-dimensional.
Then the null and the alternative models are written as
\[\begin{array}{l}
 \mbox{H}_0:\ (\beta_{\nu},\ldots,\beta_{k-1}) = (0,\ldots,0),\\
 \mbox{H}_1:\ (\beta_{\nu},\ldots,\beta_{k-1}) \neq (0,\ldots,0),
\end{array}\]
respectively.
On the other hand, if we consider significance test of a single 
additional effect (which can be a main effect or an interaction effect),
the alternative model is written in the form of
\begin{equation}
\begin{array}{l}
 \mbox{H}_1:\ (\beta_{\nu},\ldots,\beta_{\nu+m}) \neq (0,\ldots,0),
\end{array}
\label{eqn:additional}
\end{equation}
where 
$m=1$ for the case of two-level factors considered  in 
\cite{AT06}. 
For the case of three-level factors, $m$ is $2,4,8,\ldots$, depending on
the degree of freedom for the factors we consider. 
We explain this point in Section 2.2.

Depending on the hypotheses, we also specify appropriate test statistic
$T(\By)$. For example, the likelihood ratio statistic or the Pearson
goodness-of-fit statistic are frequently used. Once we specify the null
model and the test statistic, our purpose is to calculate the $p$
value. Similarly to the context of the analysis of the contingency
tables, Markov chain Monte Carlo procedure is a valuable tool,
especially when the traditional large-sample approximation is inadequate
and the exact calculation of the $p$ value is infeasible. 
To perform the Markov chain Monte Carlo procedure, 
the key notion is to calculate a Markov
basis over the sample space
\begin{equation}
 {\cal F}(X'\By^o) = \{\By\ |\ X'\By = X'\By^o,\ y_i\ \mbox{is a
 nonnegative integer for $i = 1,\ldots,k$}\},
\label{eq:t-fiber}
\end{equation}
where $\By^o$ is the observed count vector. 
Once a Markov basis is
calculated, we can construct a connected, aperiodic and reversible
Markov chain over (\ref{eq:t-fiber}).  By the Metropolis-Hastings
procedure,
the chain can be modified to have
a stationary distribution as the conditional distribution under the null
model, which is written as
\[
 f(\By\ |\ X'\By = X'\By^o) = C(X'\By^o)\prod_{i = 1}^k\frac{1}{y_i!} ,
\]
where $C(X'\By^o)$ is the normalizing constant
determined from $X'\By^o$ defined as
\[
 C(X'\By^o)^{-1} = \displaystyle\sum_{\By \in {\cal F}(X'\By^o)}\left(
\displaystyle\prod_{i = 1}^k\frac{1}{y_i!}
\right).
\]
For the definition of Markov basis 
see \cite{DS98} and for 
computational details of Markov chains see \cite{Ri87}.
In applications, it is most convenient to rely on algebraic computational
softwares such as 4ti2 (\cite{4ti2}) to derive a Markov basis.

\subsection{How to define the covariate matrix}
As we have seen in Section 2.1, it is a key formalization to express
various models by the covariate matrix $X$. The matrix $X$ is constructed
from the design matrix to reflect the main and the interaction effects
of the factors which we intend to measure. 

For the case of two-level factors, we have already considered this problem
in \cite{AT06}. In the case of two-level factors, each main effect and
interaction effect can be represented as one column of $X$. 
This is because each main and interaction effect has one degree of
freedom in the two-level case. For example of Table 1, the main effect
model of the seven factors, $\BA,\BB,\BC,\BD,\BE,\BF,\BG$ can be
represented as the $16 \times 8$ covariate matrix
by defining $\Bx_j \in \{0,1\}^{16}$ in (\ref{eqn:def-X}) as the levels
for the $j$-th factor given in Table 1.
If we intend to include, for example,  the interaction effect of $\BA \times \BB$, 
the column
\[
 (1,1,1,1,0,0,0,0,0,0,0,0,1,1,1,1)'
\]
is added to $X$, which represents the contrast of $\BA \times \BB$. 
It should be noted that Markov basis for testing the null hypothesis
depends on the model, namely the choice of various interaction effects
included in $X$.

In this paper, we consider the case of three-level designs. 
To explain three-level fractional factorial designs, first we
consider $3^p$ full factorial designs. 
Since a $3^p$ full factorial design is a special case of a multi-way
layout, we can use the notions of ANOVA model.
In this case, each main effect has two degrees of freedom since each
factor has three levels. Similarly, each two-factor interaction has
$(3-1)(3-1)=4$
degrees of freedom, three-factor interaction has $(3-1)(3-1)(3-1)=8$
degrees of freedom and so on. 
As is noted in \cite{WH00}, these sum of squares are further decomposed
into components, each with two degrees of freedom. Consider, for
example, two-factor interaction $\BA \times \BB$. We write the levels
of the factors $\BA,\BB,\BC,\ldots$ as $a,b,c\ldots \in\{0,1,2\}$ hereafter.
Then  
$\BA \times \BB$ interaction effect is decomposed to two components
denoted $\BA\BB$ and $\BA\BB^2$, where $\BA\BB$ represents the contrasts
satisfying
\[
 a + b = 0,1,2\ (\mbox{mod}\ 3),
\]
and $\BA\BB^2$ represents the contrasts satisfying
\begin{equation}
 a + 2b = 0,1,2\ (\mbox{mod}\ 3),
\label{eqn:contrast-AB^2}
\end{equation}
respectively. Since the contrasts compare values at three levels,
$0,1,2$ (mod\ $3$), each of $\BA\BB$ and $\BA\BB^2$ has two degrees of
freedom. We note the contrasts given by (\ref{eqn:contrast-AB^2}) are
equivalent to the contrasts given by 
\[
 2a + b = 0,1,2\ (\mbox{mod}\ 3),
\]
by relabeling the indices. Following to \cite{WH00}, 
we use the notational convention
that {\it the coefficient for the first nonzero factor is 1},
to avoid ambiguity. Similarly, $n$-factor interaction effects, which
have $2^n$ degrees of freedom, can be decomposed to $2^{n-1}$ components
with two degrees of freedom. For example, the three-factor interaction 
$\BA\times \BB\times \BC$ is decomposed to the $4$ components
\[\begin{array}{c}
 \BA\BB\BC, \BA\BB\BC^2, \BA\BB^2\BC, \BA\BB^2\BC^2
\end{array}
\]
and the four-factor interaction $\BA\times \BB\times \BC\times \BD$ is
decomposed to the $8$ components
\[\begin{array}{cccc}
 \BA\BB\BC\BD, & \BA\BB\BC\BD^2, & \BA\BB\BC^2\BD, & \BA\BB\BC^2\BD^2,\\
 \BA\BB^2\BC\BD, & \BA\BB^2\BC\BD^2, & \BA\BB^2\BC^2\BD, &
  \BA\BB^2\BC^2\BD^2.
\end{array}
\]

Now we explain how to define the covariate matrix $X$. 
For the full factorial designs, $X$ is constructed to include the main
and the interaction effects 
{\it with two columns for each component of two degrees}. For example of $3^3$ full
factorial design, the covariate matrix for the main effects model of
$\BA,\BB,\BC$ is written as
\[
 X' = \left[\begin{array}{ccccccccccccccccccccccccccc}
1 & 1 & 1 & 1 & 1 & 1 & 1 & 1 & 1 & 1 & 1 & 1 & 1 & 1 & 1 & 1 & 1 & 1 &
 1 & 1 & 1 & 1 & 1 & 1 & 1 & 1 & 1\\
1 & 1 & 1 & 1 & 1 & 1 & 1 & 1 & 1 & 0 & 0 & 0 & 0 & 0 & 0 & 0 & 0 & 0 &
 0 & 0 & 0 & 0 & 0 & 0 & 0 & 0 & 0\\
0 & 0 & 0 & 0 & 0 & 0 & 0 & 0 & 0 & 1 & 1 & 1 & 1 & 1 & 1 & 1 & 1 & 1 &
 0 & 0 & 0 & 0 & 0 & 0 & 0 & 0 & 0\\
1 & 1 & 1 & 0 & 0 & 0 & 0 & 0 & 0 & 1 & 1 & 1 & 0 & 0 & 0 & 0 & 0 & 0 &
 1 & 1 & 1 & 0 & 0 & 0 & 0 & 0 & 0\\
0 & 0 & 0 & 1 & 1 & 1 & 0 & 0 & 0 & 0 & 0 & 0 & 1 & 1 & 1 & 0 & 0 & 0 &
 0 & 0 & 0 & 1 & 1 & 1 & 0 & 0 & 0\\
1 & 0 & 0 & 1 & 0 & 0 & 1 & 0 & 0 & 1 & 0 & 0 & 1 & 0 & 0 & 1 & 0 & 0 &
 1 & 0 & 0 & 1 & 0 & 0 & 1 & 0 & 0\\
0 & 1 & 0 & 0 & 1 & 0 & 0 & 1 & 0 & 0 & 1 & 0 & 0 & 1 & 0 & 0 & 1 & 0 &
 0 & 1 & 0 & 0 & 1 & 0 & 0 & 1 & 0
\end{array}
\right].
\]
(We show the transpose of $X$ to save space hereafter.) 
Note that the first column represents the total mean effect, the second and
the third columns represent the contrasts of the main effect of $\BA$
and so on.
When we also consider 
the interaction effect $\BA\times \BB$, the following
four columns are added to $X$,
\[
\left[
\begin{array}{ccccccccccccccccccccccccccc}
1 & 1 & 1 & 0 & 0 & 0 & 0 & 0 & 0 & 0 & 0 & 0 & 0 & 0 & 0 & 1 & 1 & 1 &
 0 & 0 & 0 & 1 & 1 & 1 & 0 & 0 & 0\\
0 & 0 & 0 & 1 & 1 & 1 & 0 & 0 & 0 & 1 & 1 & 1 & 0 & 0 & 0 & 0 & 0 & 0 &
 0 & 0 & 0 & 0 & 0 & 0 & 1 & 1 & 1\\
1 & 1 & 1 & 0 & 0 & 0 & 0 & 0 & 0 & 0 & 0 & 0 & 1 & 1 & 1 & 0 & 0 & 0 &
 0 & 0 & 0 & 0 & 0 & 0 & 1 & 1 & 1\\
 0 & 0 & 0 & 0 & 0 & 0 & 1 & 1 & 1 & 1 & 1 & 1 & 0 & 0 & 0 & 0 & 0 & 0 &
  0 & 0 & 0 & 1 & 1 & 1 & 0 & 0 & 0
\end{array}
\right]',
\]
where each pair of columns represents the contrasts of $\BA\BB$ and
$\BA\BB^2$, respectively. For the saturated model, there are $27$
columns in $X$, i.e., one column for the total mean effect, $6 (
=2\times 3)$ columns for the contrasts of the main effects of the
factors $\BA,\BB,\BC$, $12 (= 4\times 3)$ columns for the contrasts of
the two-factor interaction effects of $\BA\times \BB, \BA\times \BC,
\BB\times \BC$ and $8$ columns for the contrasts of the three-factor
interaction effect of $\BA\times \BB\times \BC$.

Next we consider the fractional factorial designs. 
Recall a $3^{4-2}$ fractional factorial design in 
Table 2 of Section 1.
In this design, since each main effect has two degrees of freedom, the
model of the main effects for all factors, 
$\BA,\BB,\BC,\BD$, is nothing but the saturated model. 
To consider the models with interaction effects, we consider the designs of
$27$ runs. 
For example, $3^{4-1}$ fractional factorial
design of resolution IV is defined by the aliasing relation $\BD = \BA\BB\BC$. 
The relation $\BD = \BA\BB\BC$ 
means that the level $d$ of the factor $D$ is determined by the relation
\[
d = a + b + c\ (\mbox{mod}\ 3),
\]
which can also be equivalently written as
\[
a + b + c + 2d = 0,1,2\ (\mbox{mod}\ 3).
\]
Therefore this aliasing relation is also written as 
$\BA\BB\BC\BD^2 = \BI$. By the similar {\it modulus 3 calculus},
we can derive all the aliasing relations as follows.
\begin{equation}
 \begin{array}{ll}
  \BI
 = \BA\BB\BC\BD^2 & \\
  \BA = \BB\BC\BD^2 = \BA\BB^2\BC^2\BD & 
  \BB = \BA\BC\BD^2 = \BA\BB^2\BC\BD^2\\
  \BC = \BA\BB\BD^2 = \BA\BB\BC^2\BD^2 & 
  \BD = \BA\BB\BC = \BA\BB\BC\BD\\
  \BA\BB = \BC\BD^2 = \BA\BB\BC^2\BD & 
  \BA\BB^2 = \BA\BC^2\BD = \BB\BC^2\BD\\
  \BA\BC = \BB\BD^2 = \BA\BB^2\BC\BD & 
  \BA\BC^2 = \BA\BB^2\BD = \BB\BC^2\BD^2\\
  \BA\BD = \BA\BB^2\BC^2 = \BB\BC\BD & 
  \BA\BD^2 = \BB\BC = \BA\BB^2\BC^2\BD^2\\
  \BB\BC^2 = \BA\BB^2\BD^2 = \BA\BC^2\BD^2 & 
  \BB\BD = \BA\BB^2\BC = \BA\BC\BD\\
  \BC\BD = \BA\BB\BC^2 = \BA\BB\BD & 
\end{array}
\label{eqn:aliasing-relations-ABCD2}
\end{equation}
From (\ref{eqn:aliasing-relations-ABCD2}), we can clarify
the models where all the effects are estimable.
For example, the model of the main effects for the factors
$\BA,\BB,\BC,\BD$ and the interaction effects $\BA\times\BB$
is estimable, since the two components of $\BA\times\BB$, $\BA\BB$
and $\BA\BB^2$, are not confounded to any main effect.
Among the model of the main effects and two two-factor interaction
effects, the model with $\BA\times \BB$ and $\BA\times \BC$ is
estimable, while the model with $\BA\times \BB$ and $\BC\times \BD$ is
not estimable since the components $\BA\BB$ and $\BC\BD^2$ are confounded.
In \cite{WH00}, main effects or components of two-factor interaction
effects are called {\it clear} if they are not confounded to any other
main effects or components of two-factor interaction effects. Moreover, a
two-factor interaction effect, say $\BA\times \BB$ is called {\it clear}
if both of its components, $\BA\BB$ and $\BA\BB^2$, are clear. 
Therefore (\ref{eqn:aliasing-relations-ABCD2}) implies that 
each of the main effect and
the components, $\BA\BB^2, \BA\BC^2, \BA\BD, \BB\BC^2, \BB\BD, \BC\BD$
are clear, while there is no
clear two-factor interaction effect. 

\begin{remark}
\label{rem:3factor-alias}
As in Remark 
\ref{rem:2factor-alias}
the aliasing relations in
(\ref{eqn:aliasing-relations-ABCD2}) can be more elegantly
described in the framework of \cite{PRW00}. 
We consider the polynomial ring
${\mathbb C}(\BA,\BB,\BC,\BD)$ in indeterminates $\BA,\BB,\BC,\BD$.
An important note here is that, when we consider polynomials in 
${\mathbb C}(\BA,\BB,\BC,\BD)$, we cannot treat two monomials as the
same even if they designate the same contrast by relabeling indices (and 
hence we cannot use the notational 
convention of \cite{WH00}). Therefore the aliasing relations
(\ref{eqn:aliasing-relations-ABCD2}) have to be more fully written as
\begin{equation}
 \begin{array}{ll}
  \BI = \BA\BB\BC\BD^2, & \\
  \BA = \BB^2\BC^2\BD = \BA^2\BB\BC\BD^2, & 
  \BA^2 = \BB\BC\BD^2 = \BA\BB^2\BC^2\BD,\\
  \multicolumn{2}{c}{\vdots} \\
  \BA\BB = \BC^2\BD = \BA^2\BB^2\BC\BD^2, & 
  \BA\BB^2 = \BA^2\BC\BD^2 = \BB\BC^2\BD,\\
  \BA^2\BB = \BA\BC^2\BD = \BB^2\BC\BD^2, &
  \BA^2\BB^2 = \BC\BD^2 = \BA\BB\BC^2\BD\ . \\
 \end{array}
\label{eqn:aliasing-relations-ABCD2-full}
\end{equation}
For example $\BA = \BB\BC\BD^2 = \BA\BB^2\BC^2\BD$ in 
(\ref{eqn:aliasing-relations-ABCD2}), where $\BA$ and $\BA^2$ are identified
as the same contrast, is split into two aliasing relations
$\BA = \BB^2\BC^2\BD = \BA^2\BB\BC\BD^2$,
$\BA^2 = \BB\BC\BD^2 = \BA\BB^2\BC^2\BD$, where $\BA$ and $\BA^2$ are
treated as different monomials.

We first consider the polynomials
\begin{equation}
\label{eq:cubeone}
\BA^3  - 1,\  \BB^3 - 1,\ \BC^3 - 1, \ \BD^3 - 1.
\end{equation}
Note that the roots of $x^3=1$ are $1,\omega, \omega^2$, where
$\omega=\cos(2\pi/3)+i\sin(2\pi/3)$ is the cube root of unity.
Therefore (\ref{eq:cubeone}) corresponds to labeling the three levels
of the factors $\BA,\dots,\BD$ as $1,\omega$ or $\omega^2$.  
In the case of two-level factors, this 
corresponds to labeling levels  as $+1$ and $-1$ (rather than $0$ and
$1$).
Then the ideal
\begin{equation}
\label{eq:3ideal}
\langle
\BA^3  - 1, \BB^3 - 1,\BC^3 - 1,\BD^3 - 1, \BD - \BA\BB\BC
\rangle
\end{equation}
determines the aliasing relations, i.e., two interaction effects are
aliased in the sense of (\ref{eqn:aliasing-relations-ABCD2-full}) 
if and only if their difference belongs to (\ref{eq:3ideal}).
For example, $\BA$ and $\BB^2\BC^2\BD$ are aliased since
\[\begin{array}{l}
\BA - \BB^2\BC^2\BD\\
\  = 
  (\BB^2\BC^2\BD-\BA)(\BA^3-1)-\BA^4\BC^3(\BB^3-1)-\BA^4(\BC^3-1)-\BA^3\BB^2\BC^2(\BD-\BA\BB\BC)\\
\ \in  
\langle
\BA^3  - 1, \BB^3 - 1,\BC^3 - 1,\BD^3 - 1, \BD - \BA\BB\BC
\rangle\ .
\end{array}
\]

Note that in Example 29 of \cite{PRW00}, three levels of a factor are
coded as $\{-1,0,1\}$ and the polynomials $\BA^3 - \BA, \dots,
\BD^3 - \BD$ are used for determining the design ideal.  These
polynomials and coding by $\{-1,0,1\}$ do not
correspond to factorial designs considered in this section.
\end{remark}

\section{Correspondence to the models for contingency tables}
In this section, we investigate relation between 
fractional factorial designs with $3^{p-q}$ runs and contingency
tables. Since Markov bases have been mainly considered in the context of
contingency tables, it is convenient to characterize the relations
from the viewpoint of hierarchical models
of contingency tables. For the $2^{p-q}$ fractional factorial designs,
we have considered this topic in \cite{AT06}. 
In this paper, we show that  many interesting indispensable
fibers  with three elements  appear from the three-level designs.

\subsection{Models for the full factorial designs} 
First we consider $3^p$ full factorial design and prepare a fundamental fact. 
Our idea is to index  observations as $\By = (y_{i_1\cdots i_p}),
\ 1\leq i_1,\ldots,i_p \leq 3$, instead of $\By = (y_1,\ldots,y_k)',\ k
= 3^p$, 
to investigate the correspondence to the $3^p$ contingency table.
In the case
of the $3^2$ full factorial design, for example, the contrasts for each factor
and the observation are written as follows.
\begin{center}
\begin{tabular}{cccccc}
Run & $\BA$ & $\BB$ & $\BA\BB$ & $\BA\BB^2$ & $\By$ \\ \hline
1 & 0 & 0 & 0 & 0 & $y_{11}$\\
2 & 0 & 1 & 1 & 2 & $y_{12}$\\
3 & 0 & 2 & 2 & 1 & $y_{13}$\\
4 & 1 & 0 & 1 & 1 & $y_{21}$\\
5 & 1 & 1 & 2 & 0 & $y_{22}$\\
6 & 1 & 2 & 0 & 2 & $y_{23}$\\
7 & 2 & 0 & 2 & 2 & $y_{31}$\\
8 & 2 & 1 & 0 & 1 & $y_{32}$\\
9 & 2 & 2 & 1 & 0 & $y_{33}$\\ \hline
\end{tabular}
\end{center}
In this case, we see that the sufficient statistic for the parameter
for the total mean is expressed as $y_{\cdot\cdot}$ and, under given 
$y_{\cdot\cdot}$, 
the sufficient statistic for the parameter
of the main effects of the factors $\BA$ and $\BB$ are expressed as
$y_{i\cdot}$
and $y_{\cdot j}$, respectively. 
Moreover, it is seen that adding contrasts
for $\BA\BB$ and $\BA\BB^2$ yields the saturated model. Note that this relation
also holds for higher dimensional contingency tables, which we summarize
in the following.
We write the controllable factors as $\BA_1,\BA_2,\BA_3,\ldots$ instead 
of $\BA,\BB,\BC\ldots$ here. We also use the notation of $D$-marginal in the 
$p$-dimensional contingency tables for $D \subset \{1,\ldots,p\}$ here. 
For example, $\{1\}$-marginal, $\{2\}$-marginal, $\{3\}$-marginal
of $\By = (y_{ijk})$ are the one-dimensional tables $\{y_{i\cdot\cdot}\}$, 
$\{y_{\cdot j\cdot}\}$, $\{y_{\cdot\cdot k}\}$, respectively, and 
$\{1,2\}$-marginal, $\{1,3\}$-marginal, $\{2,3\}$-marginal
of $\By = (y_{ijk})$ are the two-dimensional tables $\{y_{ij\cdot}\}$, 
$\{y_{i\cdot k}\}$, $\{y_{\cdot jk}\}$, respectively. See \cite{D03} for the
formal definition.

\begin{fact}
For $3^p$ full factorial design, write observations as $\By =
 (y_{i_1\cdots i_p})$.
Then the necessary and the sufficient condition that the 
$\{i_1,\ldots,i_n\}$-marginal $n$-dimensional table $(n \leq p)$ is
 uniquely determined 
from $X'\By$ is that
the covariate matrix $X$ includes
the contrasts for all the components of $m$-factor interaction effects
$\BA_{j_1}\times \BA_{j_2}\times\cdots\times \BA_{j_m}$ for all
$\{j_1,\ldots,j_m\} \subset \{i_1,\ldots,i_n\}, m \leq n$.
\end{fact}

This fact is easily proved as follows.
The saturated model for the $3^n$ full factorial
design is expressed as the contrast for the total mean, 
$2\times n$ contrasts for the main effects, 
$2^{m}\times {n\choose{m}}$ contrasts for the $m$-factor interaction effects
for $m = 2,\ldots,n$, since they are linearly independent and
\[
1 + 2n + \sum_{m = 2}^{n}2^{m}{n\choose{m}} = (1+2)^n = 3^n .
\]

\subsection{Models for the fractional factorial designs} 
Fact 3.1 states that hierarchical models 
for the controllable
factors in the $3^p$ full factorial design corresponds to the
hierarchical models
for the $3^p$ contingency table completely. 
On the other hand, hierarchical models 
for the controllable factors in the $3^{p-q}$ fractional factorial design 
do not correspond to the hierarchical models 
for the $3^p$ contingency table in general. This is because $X$ contains only
part of the contrasts of interaction elements in the case of fractional
factorial designs.
Consequently, many interesting structures appear in considering the
Markov basis for
the fractional factorial designs. 

As a simplest example, we first consider a design with $9$ runs for
three controllable
factors, i.e., $3^{3-1}$ fractional factorial design. Write three
controllable factors
as $\BA,\BB,\BC$, and define $\BC = \BA\BB$. The design is represented
as Table 2 by ignoring the factor $\BD$. In this design, the covariate
matrix for the 
main effects model of $\BA,\BB,\BC$
is defined as
\[
X' = \left[
\begin{array}{ccccccccc}
1 & 1 & 1 & 1 & 1 & 1 & 1 & 1 & 1\\
1 & 1 & 1 & 0 & 0 & 0 & 0 & 0 & 0\\
0 & 0 & 0 & 1 & 1 & 1 & 0 & 0 & 0\\
1 & 0 & 0 & 1 & 0 & 0 & 1 & 0 & 0\\
0 & 1 & 0 & 0 & 1 & 0 & 0 & 1 & 0\\
1 & 0 & 0 & 0 & 0 & 1 & 0 & 1 & 0\\
0 & 1 & 0 & 1 & 0 & 0 & 0 & 0 & 1\\
\end{array}
\right].
\]
To investigate the structure of the fiber, write the observation as a frequency of the
$3\times 3$ contingency table, $y_{11},\ldots,y_{33}$. Then the fiber is the set of 
tables with the same 
row sums $\{y_{i\cdot}\}$, column sums $\{y_{\cdot j}\}$ and the contrast displayed as
\[
\begin{array}{|ccc|}\hline
0 & 1 & 2\\
1 & 2 & 0\\
2 & 0 & 1\\ \hline
\end{array}\ .
\]
To construct a minimal Markov basis, we see that the moves to connect 
the following three-elements fiber are sufficient
\[
\left\{\ 
\begin{array}{|ccc|}\hline
1 & 0 & 0\\
0 & 1 & 0\\
0 & 0 & 1\\ \hline
\end{array}\ ,\ \  
\begin{array}{|ccc|}\hline
0 & 1 & 0\\
0 & 0 & 1\\
1 & 0 & 0\\ \hline
\end{array}\ ,\ \  
\begin{array}{|ccc|}\hline
0 & 0 & 1\\ 
1 & 0 & 0\\
0 & 1 & 0\\ \hline
\end{array}\ 
\right\}\ .
\]
Therefore any two moves from the  set 
\[
\left\{\ 
\begin{array}{|ccc|}\hline
+1 & -1 &  0\\
 0 & +1 & -1\\
-1 &  0 & +1\\ \hline
\end{array}\ ,\ \  
\begin{array}{|ccc|}\hline
+1 &  0 & -1\\
-1 & +1 &  0\\
 0 & -1 & +1\\ \hline
\end{array}\ ,\ \  
\begin{array}{|ccc|}\hline
0 & +1 & -1\\ 
-1 & 0 & +1\\
+1 & -1 & 0\\ \hline
\end{array}\ 
\right\}
\]
is a minimal Markov basis. In the following, to save the space, we use
a binomial representation. For example, the above three moves are written as
\[
y_{11}y_{22}y_{33}-y_{12}y_{23}y_{31},\ 
y_{11}y_{22}y_{33}-y_{13}y_{21}y_{32},\ 
y_{12}y_{23}y_{31}-y_{13}y_{21}y_{32}.
\]

In this paper, we consider three types of 
fractional factorial designs with $27$ runs, which are important for
practical applications. We investigate
the relations between 
various models for the fractional factorial designs and the $3\times
3\times 3$ contingency table.
In the context of the Markov basis for the contingency tables, Markov
basis for the 
$3\times 3\times 3$ contingency tables have been investigated by many
researchers, 
especially for the no three-factor interaction model by \cite{AT03}.
In the following, we investigate Markov bases for some models, especially
we are concerned about their {\it minimality}, {\it unique minimality} and 
{\it indispensability} of their elements.
These concepts are presented in \cite{TA04} and \cite{ATY05}.
In this paper, we define that a Markov basis is minimal if no proper 
subset of it is a Markov basis. A minimal Markov basis is unique if there
is only one minimal Markov basis except for sign changes of their elements.
An element of a Markov basis is represented as a binomial. We call it a move
following our previous papers.
A move $\Bz$ is indispensable if $\Bz$ or $-\Bz$ belongs to every Markov
basis.

\paragraph*{$3_{IV}^{4-1}$ fractional factorial design defined from $\BD
    = \BA\BB\BC$}  \quad
In the case of four controllable factors for design with $27$ runs, 
we have a resolution IV design by setting $\BD = \BA\BB\BC$.
As seen in Section 2.2, 
all main effects are clear, whereas all the two-factor interactions are
not clear
in this design.

For the main effect model in this design, the sufficient statistic is
written as
\[
\{y_{i\cdot\cdot}\},\ \{y_{\cdot j\cdot}\},\ \{y_{\cdot\cdot k}\}
\]
and the contrasts of $\BA\BB\BC$,
\[
\begin{array}{l}
 y_{111}+y_{123}+y_{132}+y_{213}+y_{222}+y_{231}+y_{312}+y_{321}+y_{333},\\
 y_{112}+y_{121}+y_{133}+y_{211}+y_{223}+y_{232}+y_{313}+y_{322}+y_{331},\\
 y_{113}+y_{122}+y_{131}+y_{212}+y_{221}+y_{233}+y_{311}+y_{323}+y_{332}.
\end{array}
\]
By calculation by 4ti2, we see that
the minimal Markov basis for this model consists of $54$ degree $2$ moves
and $24$ degree $3$ moves. All the elements of the same degrees are on the same
orbit (see \cite{AT05b},\cite{AT07}). 
The elements of degree $2$ connect three-elements
fibers such as
\begin{equation}
\{ y_{112}y_{221},\ y_{121}y_{212},\ y_{122}y_{211} \}
\label{eqn:fiber1}
\end{equation}
into a tree, and
the elements of degree $3$ connect three-elements
fibers such as
\begin{equation}
\{ y_{111}y_{122}y_{133},\ y_{112}y_{123}y_{131},\ y_{113}y_{121}y_{132} \}
\label{eqn:fiber2}
\end{equation}
into a tree.
For the fiber (\ref{eqn:fiber1}), for example, two moves such as
\[
y_{121}y_{212} - y_{112}y_{221},\ y_{122}y_{211} - y_{112}y_{221} 
\]
are needed for a Markov basis. See \cite{TA04} for detail on the structure of 
a minimal Markov basis.

Considering the aliasing relations given by
(\ref{eqn:aliasing-relations-ABCD2}),
we can consider models with interaction effects.
We see by performing 4ti2 that the structures of the minimal Markov bases 
for each model are given as follows.
\begin{itemize}
\item For the model of the main effects and the interaction effect
      $\BA\times\BB$,
$27$ indispensable moves of degree $2$ such as $y_{113}y_{321}-y_{111}y_{323}$
and $54$ dispensable moves of degree $3$ constitute
a minimal Markov basis. The degree $3$ elements are on two orbits, one connects
$9$ three-elements fibers such as (\ref{eqn:fiber2}) and the other connects
$18$ three-elements fibers such as
$\{
y_{111}y_{133}y_{212},\ y_{112}y_{131}y_{213},\ y_{113}y_{132}y_{211}
\}$. 
\item For the model of the main effects and the interaction effects
      $\BA\times\BB,\BA\times\BC$,
$6$ dispensable moves of degree $3$, $81$ indispensable moves of degree
      $4$ such as
$$
y_{112}y_{121}y_{213}y_{221}-y_{111}y_{122}y_{211}y_{223}
$$ 
and
$171$ indispensable moves of degree $6$, $63$ moves such as
\[
y_{112}y_{121}y_{133}y_{213}y_{222}y_{231} -
      y_{111}y_{123}y_{132}y_{211}y_{223}y_{232}
\]
and $108$ moves such as
\[
y_{112}y_{121}y_{213}y_{231}y_{311}y_{323} -
      y_{111}y_{122}y_{211}y_{233}y_{313}y_{321},
\]
constitute a minimal Markov basis. The degree $3$ elements connect
      three-elements fibers
such as (\ref{eqn:fiber2}).
\item For the model of the main effects and the interaction effects
      $\BA\times\BB,\BA\times\BC,\BB\times\BC$,
$27$ indispensable moves of degree $6$ such as
\[
y_{113}y_{121}y_{132}y_{211}y_{222}y_{233} -
      y_{111}y_{122}y_{133}y_{213}y_{221}y_{232}
\]
and $27$ indispensable moves of degree $8$ such as
\[
y_{111}^2y_{122}y_{133}y_{212}y_{221}y_{313}y_{331} -
      y_{112}y_{113}y_{121}y_{131}y_{211}y_{222}y_{311}y_{333}
\]
constitute a unique minimal Markov basis. 

\item For the model of the main effect and the interaction effects
      $\BA\times\BB,\BA\times\BC,\BA\times\BD$,
$6$ dispensable moves of degree $3$ constitute a minimal Markov basis,
      which connect three-elements
fibers such as (\ref{eqn:fiber2}).
\end{itemize}

\paragraph*{$3_{III}^{5-2}$ fractional factorial design defined from
    $\BD = \BA\BB, \BE = \BA\BB^2\BC$} \quad
In the case of five controllable factors for designs with $27$ runs, 
the contrasts for the two main factors are allocated by two aliasing
relations. 

In this paper, we consider two designs from Table 5A.2 of \cite{WH00}. 
First we consider the $3_{III}^{5-2}$ fractional factorial design
defined from $\BD = \BA\BB, 
\BE = \BA\BB^2\BC$.

For this design, we can consider the following nine distinct
hierarchical models (except for the
saturated model). Minimal Markov bases for these models are calculated
by 4ti2 as follows.
\begin{itemize}
\item For the model of the main effects of $\BA,\BB,\BC,\BD,\BE$, 
$27$ indispensable moves of degree 2 such as
      $y_{112}y_{221}-y_{111}y_{222}$, $56$ dispensable moves of degree
      $3$, $54$ indispensable moves of degree $4$ such as
\[
y_{112}y_{121}y_{231}y_{312}-y_{111}y_{131}y_{212}y_{322} 
\]
and $9$ indispensable moves of degree $6$ such as
\[
y_{122}y_{131}y_{211}y_{232}y_{312}y_{321} - y_{111}y_{112}y_{222}^2y_{321}^2
\]
constitute a minimal Markov basis.
The degree $3$ elements are in $3$ orbits, which connects three types of
      three-elements fibers, i.e., 
\[\begin{array}{c}
18\ \mbox{moves for}\ 9\ \mbox{fibers such as}\  
\{y_{111}y_{123}y_{132},\ y_{113}y_{122}y_{131},\ y_{112}y_{121}y_{133}\},\\
36\ \mbox{moves for}\ 18\ \mbox{fibers such as}\  
\{y_{111}y_{123}y_{212},\ y_{113}y_{122}y_{211},\
y_{112}y_{121}y_{213}\}\ \mbox{and}\\
2\ \mbox{moves for the fiber}\ 
\{y_{112}y_{223}y_{331},\ y_{131}y_{212}y_{323},\ y_{121}y_{232}y_{313}\}.
\end{array}
\]

\item For the model of the main effects and the interaction effect
      $\BA\times\BC$, 
$18$ dispensable moves of degree $3$, $162$ indispensable moves of
      degree $4$ such as
\[
y_{112}y_{121}y_{213}y_{221}-y_{111}y_{122}y_{211}y_{223},
\]
$135$ indispensable moves of degree $5$ such as
\[\begin{array}{c}
 81\ \mbox{moves of the form}\ y_{112}y_{113}y_{121}y_{221}y_{331} -
  y_{111}y_{122}y_{123}y_{231}y_{311}\ \mbox{and}\\
 54\ \mbox{moves of the form}\ y_{112}^2y_{121}y_{221}y_{331} -
  y_{111}y_{122}y_{132}y_{211}y_{321}
\end{array}
\]
and $54$ indispensable moves of degree $6$ such as
\begin{equation}
y_{111}y_{122}y_{133}y_{211}y_{222}y_{233} -
      y_{112}y_{123}y_{131}y_{213}y_{221}y_{232}
\label{eqn:deg-6-indispensable-move}
\end{equation}
constitute a minimal Markov basis.
The degree $3$ elements connect three-elements fibers such as
\begin{equation}
 \{y_{111}y_{123}y_{132},\ y_{112}y_{121}y_{133},\ y_{113}y_{122}y_{131}\}
\label{eqn:fiber3}
\end{equation}

\item For the model of the main effects and the interaction effect
      $\BC\times\BE$, 
$27$ indispensable moves of degree $2$ such as $y_{112}y_{221}-y_{111}y_{222}$
constitute a unique minimal Markov basis.

\item For the model of the main effects and the interaction effects
      $\BA\times\BC,\BA\times\BE$, 
$6$ dispensable moves of degree $3$ and $81$ indispensable moves of
      degree $6$ such as
\[
 y_{122}y_{131}y_{211}y_{232}y_{312}y_{321} -
      y_{111}y_{112}y_{221}y_{222}y_{331}y_{332}
\]
constitute a minimal Markov basis.
The degree $3$ elements connect three-elements fibers such as
      (\ref{eqn:fiber3}).
\item For the model of the main effects and the interaction effects
      $\BA\times\BC,\BB\times\BC$, 
$27$ indispensable moves of degree $4$ such as
\[
y_{112}y_{121}y_{232}y_{311} - y_{111}y_{132}y_{212}y_{321}
\]
and $54$ indispensable moves of degree $6$ such as
\[
y_{112}y_{121}y_{133}y_{211}y_{223}y_{232} -
      y_{111}y_{123}y_{132}y_{212}y_{221}y_{233} 
\]
constitute a unique minimal Markov basis.

\item For the model of the main effects and the interaction effects
      $\BA\times\BC,\BC\times\BE$, 
$27$ indispensable moves of degree $4$ such as
\[
y_{111}y_{132}y_{211}y_{222} - y_{112}y_{131}y_{212}y_{221}
\]
and $54$ indispensable moves of degree $6$ such as
      (\ref{eqn:deg-6-indispensable-move})
constitute a unique minimal Markov basis.
\item For the model of the main effects and the interaction effects
      $\BA\times\BC,\BA\times\BE,\BC\times\BE$, 
$9$ indispensable moves of degree $6$ such as
\[
y_{113}y_{122}y_{131}y_{212}y_{221}y_{233} -
      y_{111}y_{123}y_{132}y_{211}y_{223}y_{232} 
\]
constitute a unique minimal Markov basis.
\item For the model of the main effects and the interaction effects
      $\BA\times\BC,\BB\times\BC,\BC\times\BD$, 
$9$ indispensable moves of degree $6$ such as
\[
y_{122}y_{131}y_{211}y_{232}y_{312}y_{321} -
      y_{111}y_{112}y_{221}y_{222}y_{331}y_{332}
\]
constitute a unique minimal Markov basis.
\item For the model of the main effects and the interaction effects
      $\BA\times\BC,\BB\times\BC,\BC\times\BE$, 
$9$ indispensable moves of degree $6$ such as
\[
y_{112}y_{121}y_{211}y_{232}y_{322}y_{331} -
      y_{111}y_{122}y_{212}y_{231}y_{321}y_{332}
\]
constitute a unique minimal Markov basis.
\end{itemize}

\paragraph*{$3_{III}^{5-2}$ fractional factorial design defined from
    $\BD = \BA\BB, \BE = \BA\BB^2$} \quad
Next we consider $3_{III}^{5-2}$ fractional factorial design defined
from $\BD = \BA\BB, \BE = \BA\BB^2$.
For this design, we can consider the following four distinct
hierarchical models (except for the
saturated model). Minimal Markov bases for these models are calculated
by 4ti2 as follows.
\begin{itemize}
\item For the model of the main effects of $\BA,\BB,\BC,\BD,\BE$, 
$108$ indispensable moves of degree $2$ such as $y_{112}y_{121} -
      y_{111}y_{122}$ 
constitute a unique minimal Markov basis.
\item For the model of the main effects and the interaction effect
      $\BA\times\BC$, 
$27$ indispensable moves of degree $2$ such as $y_{112}y_{121} -
      y_{111}y_{122}$ 
constitute a unique minimal Markov basis.
\item For the model of the main effects and the interaction effects
      $\BA\times\BC,\BB\times\BC$, 
$27$ indispensable moves of degree $4$ such as 
\[
y_{112}y_{121}y_{211}y_{222} - y_{111}y_{122}y_{212}y_{221} 
\]
and $54$ indispensable moves of degree $6$ such as 
\[
y_{112}y_{121}y_{133}y_{211}y_{223}y_{232} -
      y_{111}y_{123}y_{132}y_{212}y_{221}y_{233}
\]
constitute a unique minimal Markov basis.
\item For the model of the main effects and the interaction effects
      $\BA\times\BC,\BB\times\BC,\BC\times\BD$, 
$9$ indispensable moves of degree $6$ such as 
\[
y_{111}y_{132}y_{212}y_{221}y_{322}y_{331} -
      y_{112}y_{131}y_{211}y_{222}y_{321}y_{332}
\]
constitute a unique minimal Markov basis.
\end{itemize}

\section{Discussion}
In this paper, we investigate a Markov basis arising from the fractional
factorial designs with three-level factors. As noted in Section 1, the notion
of a Markov basis is one of the fundamental key words in the first work
of the computational algebraic statistics. Moreover, the designed
experiment is also one of the areas in statistics where the theory of
the Gr\"obner basis found applications. Since we give another
application of the theory of the Gr\"obner basis to the designed
experiments, this paper relates to both of the works by Diaconis and
Sturmfels (\cite{DS98}) and Pistone and Wynn (\cite{PW96}). 

Though we suppose that the observations are counts in Section 2, our
arguments can also be applied to the case that the observations are the
ratio of counts. In this case, we consider the logistic link function
instead of the logit link, and investigate the relation between
$3^{p-q}$ fractional factorial designs to the $3^{p-q+1}$ contingency
tables. See \cite{AT06} for the two-level case. 

One of the interesting observations of this paper is that many
three-elements fibers arise in considering minimal Markov bases. In
fact, in the examples considered in Section 3.2, all the dispensable
moves of minimal
Markov bases are needed for connecting three-elements fibers, where each
element of the fibers does not share supports in each other. This shows
that the every positive and the negative part of the dispensable moves
is a indispensable. See notion of the {\it indispensable monomial} in
\cite{ATY05}. 

It is of great interest to clarify relationships between our
approach and the works by Pistone, Riccomagno and Wynn \cite{PRW00}.
In \cite{PRW00}, designs are defined as the set of points (i.e., the
affine variety), and the set of polynomials vanishing at these
points (i.e., the design ideal) are considered. They calculate the
Gr\"obner basis of the design ideal, which is used to specify the
identifiable models or confounding relations.
In Section 2 we explained  that the aliasing relations 
for fractional factorial designs specified 
in the classical notation can be more elegantly described in the
framework of \cite{PRW00}.  It is important to study whether 
a closer connection can be established 
between a design ideal and the Markov basis (toric
ideal). It should be noted, however that  a
Markov basis depends on the covariate matrix $X$, 
which incorporates the  statistical model  we aim to test, 
whereas the Gr\"obner basis depends only on 
the design points under a given term order.

Finally as suggested by a referee, it may be valuable to consider 
relations between the arguments of this paper and designs other than
fractional factorial designs, such as
the Plackett-Burman designs or the balanced incomplete block designs.  
These topics are left to  future works.

\bibliographystyle{plain}

\begin{thebibliography}{10}

\bibitem{4ti2}
4ti2 team.
4ti2 -- a software package for algebraic, geometric and combinatorial
  problems on linear spaces.
Available at www.4ti2.de.

%

\bibitem{AHOT07}
Aoki, S., Hibi, T., Ohsugi, H. and Takemura, A. (2007).
Markov basis and Gr\"obner basis of Segre-Veronese configuration for
	testing independence in group-wise selections.
{\it METR Technical Report, 2007-21}, arXiv.org, 0704.1074.

\bibitem{AT03}
Aoki, S. and Takemura, A. (2003).
Minimal basis for a connected {M}arkov chain over {$3\times 3\times
  K$} contingency tables with fixed two-dimensional marginals.
{\it Australian and New Zealand Journal of Statistics}, {\bf 45}, 229--249.

\bibitem{AT05a}
Aoki, S. and Takemura, A. (2005).
Markov chain Monte Carlo exact tests for incomplete two-way
  contingency tables.
{\it Journal of Statistical Computation and Simulation}, {\bf 75}, 787--812.

\bibitem{AT06}
Aoki, S. and Takemura, A. (2006).
Markov chain Monte Carlo tests for designed experiments.
{\it METR Technical Report, 2006-56}. arXiv.org, math.ST/0611463.


\bibitem{AT07}
Aoki, S. and Takemura, A. (2007a).
Minimal invariant Markov basis for sampling contingency tables with
  fixed marginals.
{\it Annals of the Institute of Statistical Mathematics}, To appear.

\bibitem{AT05b}
Aoki, S. and Takemura, A. (2007b).
The largest group of invariance for Markov bases and toric ideals.
{\it Journal of Symbolic Computation}, in press. 
doi: 10.1016/j.jsc.2007.11.002.

\bibitem{ATY05}
Aoki, S., Takemura, A. and Yoshida, R. (2007c).
Indispensable monomials of toric ideals and Markov bases. 
Journal of Symbolic Computation, in press. doi: 10.1016/j.jsc.2007.07.012.

\bibitem{C93}
Condra, L. W. (1993).
{\it Reliability Improvement with Design of Experiments}.
Marcel Dekker, New York, NY.




\bibitem{DS98}
Diaconis, P. and Sturmfels, B. (1998). 
Algebraic algorithms for sampling from conditional distributions.
{\it Ann. Statist.}, {\bf 26}, 363--397.


\bibitem{D03}
Dobra, A. (2003).
Markov bases for decomposable graphical models.
{\it Bernoulli}, {\bf 9}, 1093--1108.


\bibitem{GPR03}
Galetto, F., Pistone, G. and Rogantin, M. P. (2003).
Confounding revisited with commutative computational algebra.
{\it Journal of Statistical Planning and Inference}, {\bf 117}, 345--363.


\bibitem{HN97}
Hamada, M. and Nelder, J. A. (1997).
Generalized linear models for quality-improvement experiments.
{\it Journal of Quality Technology}, {\bf 29}, 292--304.

\bibitem{HAT06}
Hara, H., Aoki, S. and Takemura, A. (2006).
Fibers of sample size two of hierarchical models and Markov bases of
	decomposable models for contingency tables. 
{\it METR Technical Report, 2006-66}. arXiv.org, math.ST/0701429.







\bibitem{MN89}
McCullagh, P. and Nelder, J. A.
{\it Generalized Linear Models, 2nd ed.}
Chapman \& Hall, London.




\bibitem{PRW00}
Pistone, G., Riccomagno, E. and Wynn, H. P. (2000).
{\it Algebraic Statistics, Computational Commutative Algebra in
  Statistics}. Chapman \& Hall.

\bibitem{PR07}
Pistone, G. and Rogantin, M. P. (2007).
Indicator function and complex coding for mixed fractional factorial
	designs. {\it Journal of Statistical Planning and Inference}.

\bibitem{PW96}
Pistone, G. and Wynn, H. P. (1996).
Generalised confounding with Gr\"obner bases.
{\it Biometrika}, {\bf 83}, 653--666.



\bibitem{Ri87}
Ripley, B. D.
{\it Stochastic Simulation}.
Wiley Series in Probability and Mathematical Statistics: Applied
  Probability and Statistics. John Wiley \& Sons Inc., New York.

\bibitem{RR98}
Robbiano, L. and Rogantin, M. P. (1998).
Full factorial designs and distracted fractions.
In {\it Gr\"obner bases and applications (Linz, 1998)}, volume 251 of
{\it London Math. Soc. Lecture Note Ser.}, pages 473--482. Cambridge Univ.
Press, Cambridge.





\bibitem{TA04}
Takemura, A. and Aoki, S. (2004).
Some characterizations of minimal Markov basis for sampling from
  discrete conditional distributions.
{\it Annals of the Institute of Statistical Mathematics}, {\bf 56},
	1--17.

\bibitem{TA05}
Takemura, A. and Aoki, S. (2005).
Distance reducing Markov bases for sampling from a discrete sample
  space.
{\it Bernoulli}, {\bf 11}, 793--813.

\bibitem{WH00}
Wu, C. F. J. and Hamada, M.
{\it Experiments: Planning, analysis, and parameter design
  optimization}. (2000)
Wiley Series in Probability and Statistics: Texts and References
  Section. John Wiley \& Sons Inc., New York.
A Wiley-Interscience Publication.

\end{thebibliography}

\end{document}